# Bring Your Own Devices Classroom: Issues of Digital Divides in Teaching and Learning Contexts

**Janak Adhikari**
School of Business
Bay of Plenty Polytechnic
Tauranga, New Zealand
Janak.Adhikari@boppoly.ac.nz

**Anuradha Mathrani**
School of Engineering and Advanced Technology
Massey University
Auckland, New Zealand
A.S.Mathrani@massey.ac.nz

**David Parsons**
The Mind Lab
Unitec Institute of Technology
Auckland, New Zealand
David@themindlab.com

## Abstract

Technology mediated learning provides potentially valuable resources for learners' academic and social development. However, according to recent researches, as the adoption stages of ICTs advance there arises further levels of digital divides in terms of equity of information literacy and learning outcomes. For the last three years we have been working with one of the earliest secondary school in New Zealand to introduce a Bring Your Own Device (BYOD) policy. Our research has included a number of methods, including surveys, interviews and classroom observations. In this paper we present the findings from the investigation into BYOD project, which offers new insights into the digital divide issues in the context of technology mediated learning. Teaching and learning practices are evolving continually across formal and informal spaces, and this study informs us how the BYOD policy has influenced existing divides in the learning process.

## 1. Introduction

With the increased diffusion of digital technologies into almost every aspect of human lives, the need for appropriate digital and information literacy for every individual is on the rise. Digital skills are now considered as the third most important life skill alongside numeracy and literacy (DfES, 2003; Johnson, Levine, Smith, & Stone, 2010).

Recent trends in formal education emphasize integration of digital learning media into existing pedagogies to transform teaching and learning (Anderson, 2009; Prestridge, 2007). Introducing ICT provides potentially valuable resources for learners' academic and social development, such as new learning activities, improved collaboration mediums, novel assessment models, and curriculum changes which introduce more visual stimulants in the learning environment (Demiraslan & Usluel, 2008). The results from early digital opportunities projects[1] in New Zealand indicated that integration of ICTs into learning might end up contributing nothing more than an effort to facilitate material access to ICTs (Rivers & Rivers, 2004). Despite the potential of innovative ICTs to improve the learning outcomes for every learner, evaluation of the projects indicate that integration of ICTs into the learning process is challenging and any such initiatives may even end up accentuating existing digital divides (Parr & Ward, 2004; Rivers & Rivers, 2004; Winter, 2004).

In response to this evaluation report, an ICT strategic framework for education has been developed in 2006 by taking account of the lessons learned from previous projects. The goal of this framework was to develop a more learner-centred service culture where education agencies and organizations focus on the outcomes rather than the technology through improved connectivity (access to ICT infrastructure for education), content (digital content from variety of sources), and capability (skills needed to turn information into knowledge) (Ministry of Education, 2006).

---

[1] The Digital Opportunities (DigiOps) projects are joint partnerships between schools, organisations involved in ICT and the Ministry of Education in New Zealand.

For the last three years we have been working with one of the earliest secondary school in New Zealand to introduce a Bring Your Own Device (BYOD) initiative based on their recommendations for using iPad in classrooms. The school, at first, recommended iPads as a preferred device, but this has since been extended to other tablets and computing devices. Our research study employed a number of methods, including surveys, interviews and classroom observations. In this paper we present the findings, which gave us insights into the digital divide issues in the context of technology mediated learning. Our initial findings reveal that equity of access and skills are not major issues, but some of the findings strengthen the need to extend the digital divide research in learning towards additional fields of enquiry (i.e. learning outcomes divide).

## 2. The Meaning of equity

From the analysis of previous digital opportunities projects, two major limitations of integration of digital learning mediums into existing pedagogy were identified. First, during the planning and implementation of the projects, the meaning of equity was understood only as a matter of material access and digital skills. However, the outcome of the projects indicated that equity in these two aspects may be a necessary first step, but is not sufficient. To address the issue of digital divides in learning, there must also be equity in learning outcomes beyond just access and skills (Wei, Teo, Chan, & Tan, 2011). According to different researchers in this field, equity of students' learning outcomes depends on factors such as (a) the attitude and motivation of students towards technology, (b) the nature of technology usage by students and (c) students' capability of meaning making (Jones & Issroff, 2007; Van Dijk, 2006; Wei et al., 2011). A second limitation was the lack of detailed forethought by planners in understanding how learning activities and environments are affected by the introduction of ICTs. According to Salomon (1993, p. 189), *"Introduction of ICTs redefines the whole activities and interpersonal relationships inside and outside of the classroom"*. Therefore, both formal (classrooms and wider school environment) and informal (home and outside school) learning spaces should be equally taken into consideration while investigating technology mediated learning.

## 3. Theoretical framework

According to recent research, as the adoption stages of ICTs advance there arise further levels of digital divides in terms of equity of information literacy and learning outcomes (Wei et al., 2011) as shown in Figure 1.

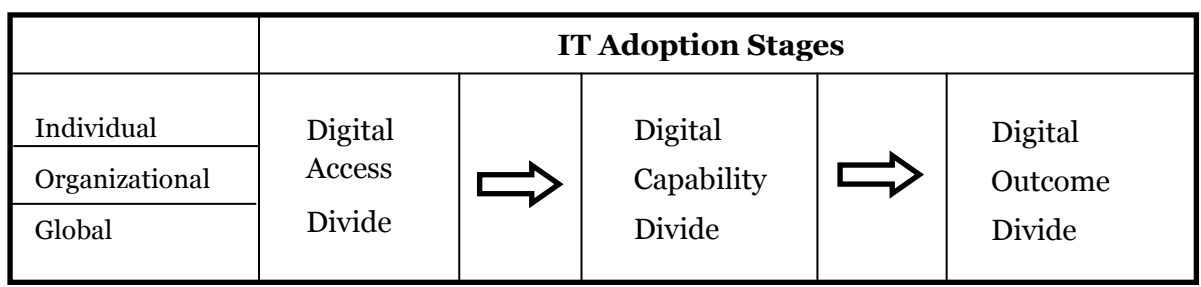

*Figure 1: Three level digital divide framework (Wei et al., 2011)*

Findings from past and current digital opportunity projects show that equitable material access to ICTs at home and school and having appropriate digital skills are necessary first steps, however, this alone is not sufficient for achieving equalised learning outcomes for every learner. There are still some unanswered questions around whether capability divide leads to outcome divide? And it is especially unclear how access to and use of technology at home may influence interactions within the school's ICT environment and vice versa, that is, how digital access divide will impact the digital capability divide and learning outcomes or the digital outcome divide. Therefore, there is a need to extend the digital divide research in the context of ICT integration in learning towards additional fields of enquiry beyond just access and skills.

The three level digital divide framework describes factors pertaining to the digital access divide to include access to and use of ICT at homes and at schools, personal attributes like gender and academic ability, and environmental conditions of homes and schools. This further influences affordances in various sources of social cognitive abilities related to individual's learning activities and computer self-efficacy levels, demonstrating digital capability divide among individuals (Wei et al., 2011). These will in turn, affect how new skills and knowledge are gained having further implications on an individual's learning outcomes leading to digital outcome divide.

For that reason, we have adopted three level digital divide framework and applied it to the context of our study. While adapting the three level digital divide framework for our study, we mapped the three levels of IT adoption stages to the three levels of digital divides in the learning process. Specifically, ICT adoption stages has access, capability and outcome divide stages, which matches with the digital access divide, digital capability divide and digital/learning outcome divide.

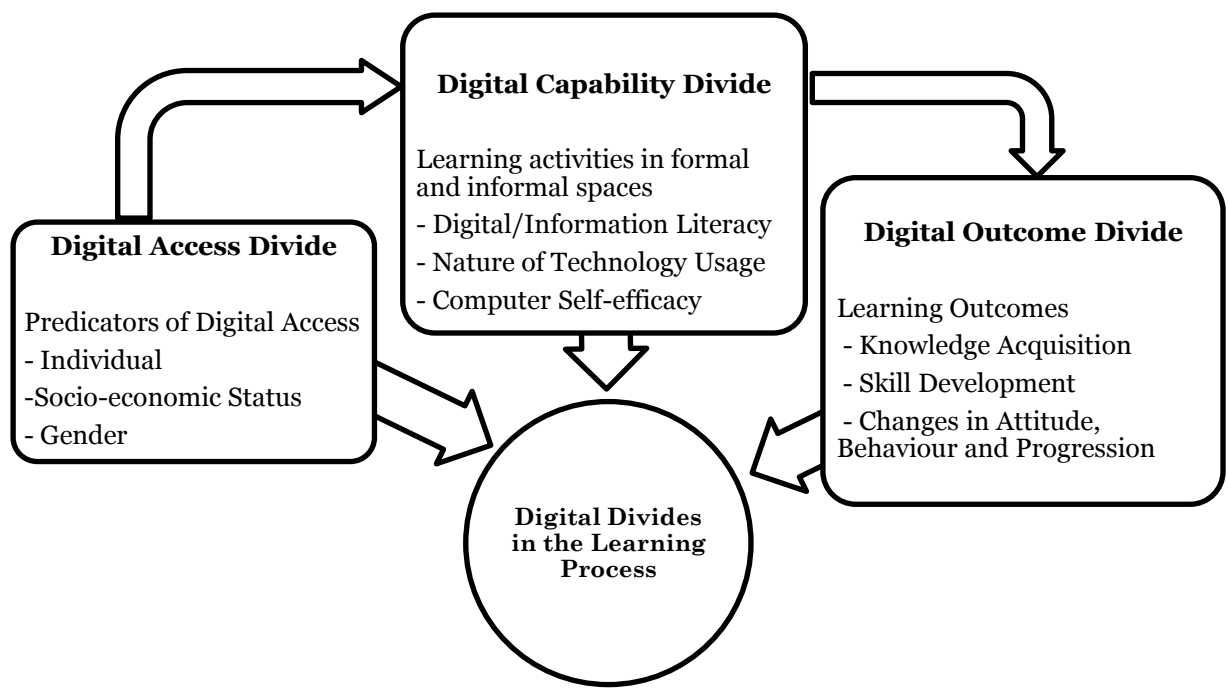

*Figure 2: Three levels of digital divide in learning applied to the context of our study.*

As illustrated in Figure 2, the digital capability divide is influenced by learning activities which occur in both formal and informal spaces. These activities can be contextualized based on digital/information literacy of learners in performing different types of computing tasks, nature of technology usage by learners ranging from familiarity to addiction, and computer self-efficacy measurements of their capabilities. Finally, digital/information literacy, computer self-efficacy and nature of technology usage are the focal constructs through which personal, behavioural and environmental factors further influence learning outcomes resulting in digital outcome divide. Accordingly, as shown in the framework, the nature of digital divide may change from one form to another over the different stages of technology adoption. Therefore, to examine digital outcome divide, various factors in first two levels of digital divides has an effect on extent of knowledge acquisition, skills development and changes in attitudes, behaviours and progression in learning.

## 4. Research objectives and methods

Drawing on the three digital divides influencing the learning process, the purpose of this research study is (a) to investigate whether and, if so, how, the introduction of BYOD initiative has changed digital divides and affected teaching and learning process, in both formal and informal learning spaces; (b) and, to evaluate the effectiveness of BYOD initiative on students' learning outcomes.

A longitudinal case study method has been employed for the study of the BYOD initiative. The case study method is particularly suited to learning in detail through an in-depth study (Dubé & Paré, 2003). Case studies are defined in various ways and a standard template does not exist, however in general, a case study examines a phenomenon in its natural setting, employing multiple methods of data collection to gather information from one or more entities. The boundaries of the phenomenon are not clearly evident at the outset of the research and no experimental control or manipulation is used (Benbasat, Goldstein, & Mead, 1987; Dubé & Paré, 2003; Yin, 2003). Our study is following a single case with repeated investigation over a period of time. According to Yin (2003), a single case design is appropriate when it represents a unique, revelatory or critical case. According to the three level digital divide framework, each of the three level of digital divide can be studied at individual, the organisational, or the country/global levels. For example, the digital access divide can be measured by computer/internet access at home (individual level), IT investment by organisation (organisational level), and national IT expenditure by countries (global/country level) (Wei et al., 2011). Our study focuses on individual learners. Therefore, we have selected just the single case which is representative of the research problem and field of enquiry we are investigating since the said case is one of the earliest adopters of BYOD in New Zealand.

## 4.1  Data Collection

Because of the nature of our investigation, we have conducted data collection in regular intervals over the last three years. Up until now, three rounds of data collection including interviews, online surveys, and classroom observations have been carried out. The online surveys and interviews with students, teachers and parents have been designed to understand and investigate equity in terms of level of access, digital skills and learning outcomes.

| Respondents | 2012 | 2013 | 2014 |
|---|---|---|---|
| Teachers | 14 | 40 | 63 |
| Parents | 4 | 71 | 50 |
| Students | 56 | 98 | 41 |

*Table 1. Numbers of respondents to each survey*

It may be noted that online surveys are limited to access of participants to one-to-one devices. To address this limitation, the school encouraged students to complete their surveys during school hours, since the school had provision for one-to-one internet-enabled devices for all students.

Apart from surveys, total of 26 one-to-one interviews have been conducted (10 students, 9 teachers and 7 parents). Also, 9 classroom observations have been conducted for target subject areas (mathematics, science, and physical education).

## 4.2  Data Analysis

Survey results were mostly quantitative in nature including some text responses. Further to quantitative analysis of the survey results across the different factors included in the three level digital divide framework (Figure 2), qualitative data has also been analysed. Interviews, classroom observations and text responses data from surveys have been coded into various categories to gain in-depth understanding of each of the themes emerging from the data. Following table 2 shows the major themes emerging from the coding of interviews, classroom observations and text responses data.

| Code No. | Coding Themes | Number Coding References |
|---|---|---|
| C1 | Students' level of access to ICTs | 16 |
| C2 | Level of digital skills and information literacy in teachers and students | 22 |
| C3 | Students' ICT usage patterns and their activities | 18 |
| C4 | Students' attitude and motivation towards ICT mediated learning | 32 |
| C5 | Challenges and issues experienced by students | 26 |

| C6 | Challenges and issues experienced by teachers | 21 |

*Table 2. Coding categories and their respective number of coding references in interview, classroom observation and text responses data related to digital divide aspect of study.*

Among the themes emerging from the coding of the qualitative data, almost all of them relate to the factors we are considering for the three level digital divide framework (Figure 2). Code C1 (students' level of access to ICTs) relates to access to ICT divide, C2 and C3 (level of digital skills and information literacy in teachers and students and ICT usage patterns) relates to learning capability divide, and rest of the themes (C4, C5 and C6) which emerged relate to the learning outcome divide. Therefore, the results are presented in the same order as the framework has described the digital divides in learning process shown in Figure 2.

## 5. Findings

This section describes the key findings based on this three level digital divide model (digital access divide, digital skills divide and learning outcomes divide). This includes how the learning process has changed over the years since the BYOD initiative was rolled out. Specifically, we are focusing on the digital divide aspect of our study for the purpose of this paper and therefore results in this papers covers data that relates to the digital divide aspect of our study including surveys and interviews with students, teachers and parents.

### 5.1　Digital access divide

One of the major issues that emerged in the preliminary analysis was potentially the large division in the classroom in terms of access to digital devices. However, the baseline data shows 100% access to digital learning devices and internet (with few exception where students had to borrow school computers for their learning needs). Despite the survey results indicating 100% access, interview responses provided in-depth insight into this issue and revealed that some students have limited access to digital technologies (at least access to internet) for their learning needs. 2 out of 9 students interviewed at the beginning of BYOD investigation, had no internet access at home even if they have a one-to-one device, and therefore expressed an inability to continue learning activities while being at home. First student says, *"I usually do not spend much time with the tablet at home because I don't have internet at home. Sometimes I can't complete my work at home because of the internet"*. Similarly another student responds, *"Well in my house we don't have dialup so I only use my tablet for the project I have downloaded. I don't have internet at home."*

Findings from data indicate that majority of limited access issues reported relates to informal learning spaces (home and outside school). Further to that, socio-economic status and geographical locations have emerged as main reasons for limited access to digital learning devices and internet, and interviews with parents' backs up the student responses. When we asked, did they think of providing one-to-one learning devices for their child; one parent says *"one-to-one devices are great for education but there needs to be equity for families that cannot afford devices"*. Another parent said asking government to provide financial support is an unrealistic expectation. However, tax breaks similar to school donations would help families in economic difficulty. One parent explained how difficult it was for some families, and this financial hardship had influenced their decision to go for cheaper non-recommended device. *"It wasn't something that was in our budget, we had to use other means to purchase this device for our daughter, it wasn't ideal as we've had to put it onto HP and with one income it has proven difficult to pay this off in the required "no interest" time frame"*.

Another issue that came up regarding digital access was the compatibility issues between different types of one-to-one devices. Interview responses from student indicate that some students have been unable to carry out their usual learning activities during classroom because of compatibility issues. On the other hand, responses from teachers confirm that most of the learning activities are designed keeping iPad in mind, and these activities may be not performed using laptop and android devices. Regarding compatibility issue, one student says *"I felt disadvantaged sometimes because I have a laptop and all the teachers talk about is apps for iPads"*. However, students and teachers were keen to find alternative ways for these situations. Also the overall survey responses do not reflect incompatibility issues to be prevalent on a larger scale in everyday learning.

Despite some degree of access, compatibility and technology issues, BYOD initiative certainly provided greater degree of access to digital learning technologies to learners and it is improving gradually.

### 5.2　Digital capability divide

According to that framework, digital/information literacy, nature of technology usage and computer self-efficacy are some of the factors that could affect the digital capability of learners. However in the context of BYOD project, findings do not provide any evidence of widening gap in digital/information literacy skills for both students and teachers. Some level of skills issues had been reported initially, but that seem to have improved in the second set of survey data.

### 5.2.1 Digital / Information Literacy

In the latest survey, overall skill levels of staff appeared to be slightly lower in the 2014 survey than in 2012 (Figure 3). However it should be noted that the 2012 staff were early adopters who volunteered to take part in the first year of the BYOD initiative. The figures for 2014 represent a larger cohort of teachers across the school. This suggests that we cannot expect the digital skills of staff overall to reach its maximum potential until the BYOD policy has been fully rolled out across all school years so that all the staff have had the opportunity to fully develop their digital skills.

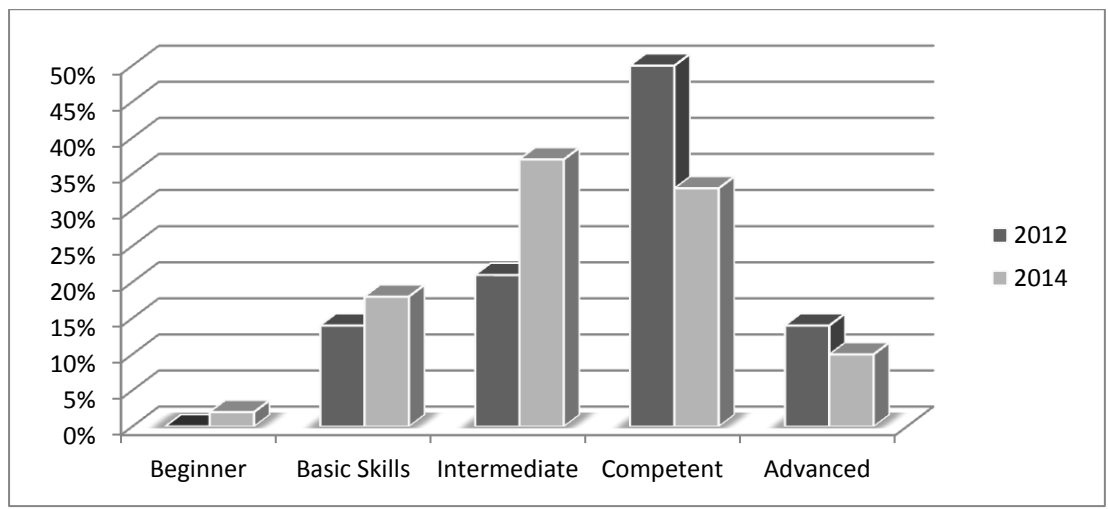

*Figure 3: Staff skill levels in digital devices and computer technology*

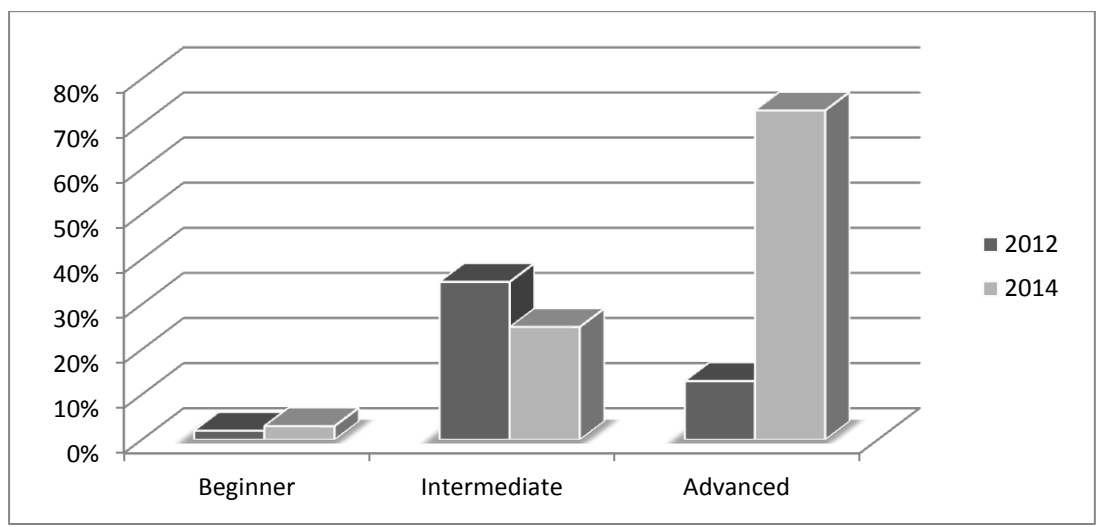

*Figure 4: Student skill levels in making meaningful use of digital devices in learning activities*

From the survey responses of students shown in Figure 4, it should be noted that we asked a somewhat different question about their levels of skill in making meaningful use of digital devices in learning. Further, the 2014 survey only had three options instead of five. Nevertheless, there is a marked increase in the perceived level of digital skills, thus we see the potential for agency has increased over time.

Learning is constantly evolving in the context of technology mediated learning environment and that started to appear in the second set of the survey responses. In recent years since the BYOD classroom, we have seen teaching and learning practices focused more on processing and applying existing information into learning, rather than creating their own content. Many students have expressed appreciation for the way they are learning and reported that BYOD classroom is clearly much more relevant and useful in today's modern society. In fact, there have been some responses that indicate this change in focus to be one of the reasons why one-to-one devices have been well received by majority of students.

Because of this shift in creating information to processing and applying information in the context of technology mediated learning, we have to extend our attention from digital skills to information literacy. The reason is, digital literacy/skills may not be the only key factor that determines the learning outcome of the students any more. Students may have very good digital skills to operate one-to-one devices but if they don't have enough skills to process and apply the information given to them, they might still struggle in achieving desired learning outcomes. Survey and interview data suggests that a significant proportion of students clearly struggle to find, process and apply information into their learning activities. Therefore, to attain the equity of learning outcomes in the changing nature of teaching and learning practices, it is necessary that we consider information literacy as the key factor to raise computer self-efficacy among students, which is one of the focal constructs in our framework.

From these results we might assume that digital/information literacy skills will increase over time, once BYOD is consistently applied across all year levels. For those who are already actively engaged in using one-to-one devices, there is certainly skill development going on. However, information literacy on the other hand is evolving as an aspect that needs more in-depth investigation in the technology mediated learning context.

### 5.2.2 Nature of technology usage by learners

Findings suggest that there is diversity in students' usage of one-to-one devices in school as well as in their everyday life. Overall their usage patterns haven't changed across the surveys. However, there is a small increase in device usage for educational purposes and that indicates the positive trend in student motivation for BYOD classrooms. However, there remains a large number (around 50%) of students reportedly spending most of their times around non-educational activities like social media/communication, games and entertainment as shown in Figure 5.

When asked what have been their major challenges, 17% of teachers responded that, keeping an eye on students during classes to prevent them from going off task remains a major challenge. Even, some of the students reported their peers going off task and classroom being disrupted because of that. Although the school has taken some measures to discourage students going off task that seems to have little or no effect. Therefore, it remains one of the challenges for school and teachers to keep learners on task. Parents in their responses, also clearly voiced their concerns regarding the unsupervised usage of devices by their child. What came out the survey results is that, parents worry about the nature of their child's device usage and potential harm to learners because of the exposure to inappropriate and damaging internet contents. One parent worried for change in their children's behaviour and social interactions says, *"Yes I constantly have to take the device off my child she seems to be constantly on it and it is a constant battle, she has lost interest in a lot of other activities"*.

However, since the initial introduction of one-to-one learning devices, there seem divides in digital literacy between students and parents. Because of this, parents also fear for the safety of their children as a result of unsupervised access to virtually unrestricted online resources. In the latest surveys, some of the parents responded as:

*"Negative effects on our family is that as everything is digital we cannot discuss what is being learnt as easily as it is not in a book to be shared but on a web site."*

*"Negative impact: they spend a huge amount of time at home on their devices. It is often very difficult for us to know whether it is school related or not. As it is a condition of them attending school we are bound to allow them access to their devices."*

*"Yes definite negative impact I have seen in our community and at home. Huge amount of social bullying and inappropriate use of the device to take photos, and send images, messages to others about others etc. Children as young as Year 7 and 8 being given complete access to the internet and everything on it getting into pornography (written and visual) and chat rooms talking to older men and women."*

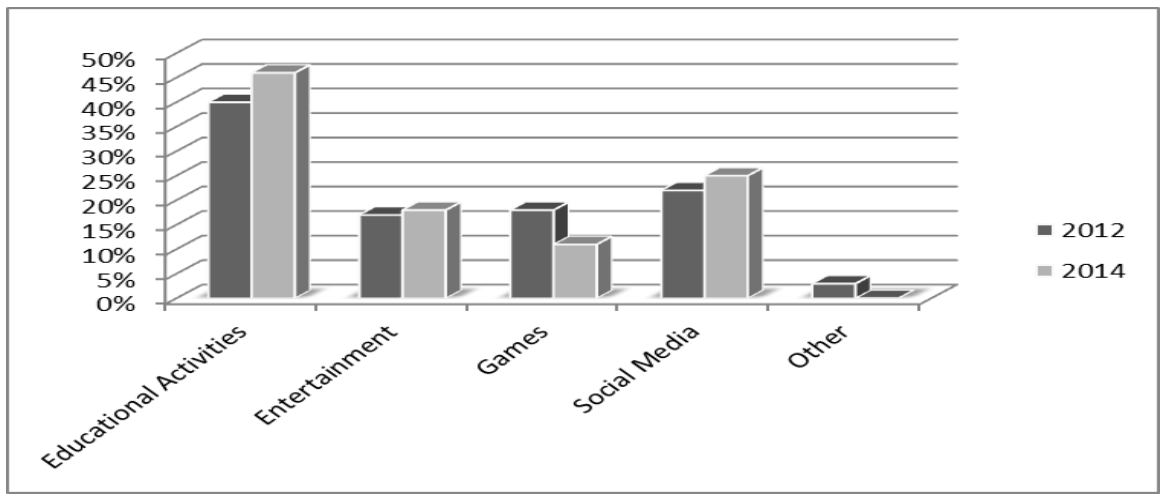

*Figure 5: Student's nature of technology usage in school and at home (self-reported)*

Lately, there have been reports of students using some of the applications and sites that are used for internet bullying in New Zealand schools. There is no report of that from the school where our research is based on, but this is clearly an alarm bell for school and parents involved in the BYOD classrooms.

### 5.3    Learning outcomes divide

Having proper digital access or skills may not be the only key factor that determines the learning outcomes of the learner any more. Students may have very good access to technologies and digital skills to operate one-to-one devices but if they don't have enough skills to process and apply the information given to them, they are still going to struggle in their learning. Therefore, our analysis in this category focuses on some of the potential factors that might have an impact on students learning outcomes:

### 5.3.1  Knowledge acquisition and skill development

Learning is constantly evolving in the context of technology mediated learning environment and that started to appear in the data in later stages. In recent years since the BYOD classroom, we have seen the teaching and learning practices focused more on processing and applying existing information into learning, rather than creating their own content. Many students have identified and appreciated the way they are learning and reported that BYOD classroom is clearly much more relevant and useful in today's modern society. In fact, there have been some responses that indicate this change in focus being one of the reasons why one-to-one devices have been well received by majority of students.

However, a small proportion of students clearly expressed their concerns regarding not being able to identify, process and apply information into their learning activities. And therefore it is much more relevant for us to investigate information literacy instead of digital skills after the change in dynamics of classroom due to BYOD classrooms.

### 5.3.2  Attitudes and motivation of learners

As expected, the majority of learners find BYOD a great idea and seemed to be happy with the changes in everyday teaching and learning as a result of that. However, some of the students expressed their unwillingness and changes in the current way of integrating devices into the use of one-to-one devices for learning activities as shown in Figure 6 below. Issues that has been raised as the reason were the compatibility with other devices, quality of internet connection at school, concerns regarding NCEA[2] exam format and lack of enough information literacy skills (not being able to find, process and apply appropriate information) for their learning needs.

---

[2] The National Certificate of Educational Achievement (NCEA) is the official secondary school qualification in New Zealand.

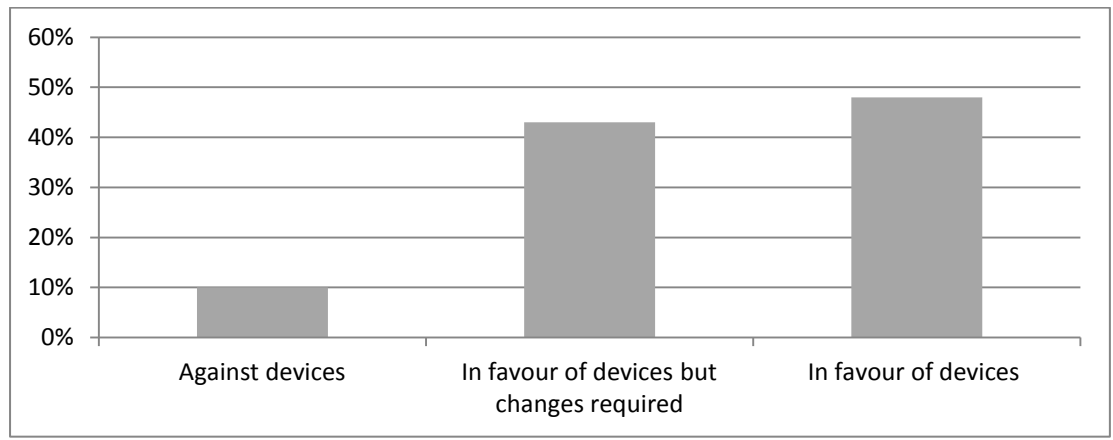

*Figure 6: Student support for digital devices in learning measured by self-reported percentages*

Students reported having some degree of concerns regarding loss in handwriting and spelling skills, since the introduction of BYOD. However, the number appeared to have increased in a latest survey. Around 50% responded that they are worried for their loss in handwriting skills, as the way they are learning is different to the way NCEA exams are conducted. Around 30% responded that they are not sure until they sit in the exams. Only 9% of students responded saying they are confident and prepared for the exam. By looking into the responses, we can conclude that a large number of students are worried for their performance in NCEA exams and this can potentially be one of the reasons behind motivational issues identified during the later stage of BYOD project.

## 6. Discussion

The integration of one-to-one learning devices has ability to transform the teaching and learning process. Findings shows that access to ICTs has improved in a phenomenal rate as a result of the BYOD policy and access to digital learning devices is not an issue in general. Findings do not provide any evidence of widening gap in digital skills for both students and teachers. Some issues in skills levels were reported initially, but that seem to be improving gradually. Learning is constantly evolving and the teaching and learning practices focused more on processing and applying existing information into learning. Because of this shift in creating information to processing and applying information in the context of technology mediated learning, we have to extend our attention from digital literacy to information literacy to further investigate the learning outcome divide.

Students' usage of one-to-one devices includes more and more use of online social media, and web 2.0 tools for everyday life as well as learning activities in formal and informal spaces. Collis & Moonen (2008) affirm the use of web 2.0 internet tools to enhance collaboration, communication, and distribution of information among its users regardless of their physical locations. By using web 2.0 social media technologies, learners can have same level of ability to carry on their learning activities from either formal or informal spaces. While this is empowering students regarding easier access to learning content and in making choices regarding preferred learning venues that suites their needs, requirements and interests, this also brings forth new avenues of divides. Parents have concerns over unsupervised access to internet while pursuing learning activities in schools and homes.

Despite progress in dealing with the changing nature of digital divides over the years, not every aspect of these divides has yet been taken into account. Ensuring equalised digital access and digital skills/literacy are necessary measures towards bridging the digital divides in the learning, but further divides still exist. In order to achieve the complete digital inclusion, objectives should be to equip learners with not only the improved access and skills to digital technologies, but the motivation and ability to think critically and creation of new knowledge which is responsive to the solution of professional and social needs (Richey, 1998).

## 7. Conclusion and Future Direction

This study has offered insights on the various aspects of digital divides in teaching and learning environment and how it has been transformed since the BYOD initiative. Data indicates more divides which are spread across different aspects and stakeholders in BYOD policy; specifically between parents and their children and between different teachers. Further to this, we will be investigating the changes in the dynamics of teaching and learning and how formal and informal learning spaces can provide different contexts in establishing technology mediated learning environments. In the next stage, a longitudinal study is planned to investigate the changes carried out in pedagogical approaches for maximizing student knowledge acquisition and skill development.

## 8. Reference

Anderson, N. (2009). *Equity and Information Communication Technology (ICT) in Education* (Vol. 6): Peter Lang Publishing Inc., New York.

Benbasat, I., Goldstein, D. K., & Mead, M. (1987). The Case Research Strategy in Studies of Information Systems. *MIS Quarterly, 11*(3), 369-386.

Collis, B., & Moonen, J. (2008). Web 2.0 Tools and Processes in Higher Education: Quality Perspectives. *Educational Media International, 45*(2), 93-106.

Demiraslan, Y., & Usluel, Y. K. (2008). ICT Integration Processes in Turkish Schools: Using Activity Theory to Study Issues and Contradictions. *Australasian Journal of Educational Technology, 24*(4), 458-474.

DfES. (2003). *21st century skills : realising our potential : individuals, employers, nation*. Retrieved from http://dera.ioe.ac.uk/4747/.

Dubé, L., & Paré, G. (2003). Rigor in Information Systems Positivist Case Research: Current Practices, Trends, and Recommendations. *MIS Quarterly, 27*(4), 597-636.

Johnson, L., Levine, A., Smith, R., & Stone, S. (2010). The 2010 Horizon Report. Austin, Texas: The New Media Consortium.

Jones, A., & Issroff, K. (2007). Motivation and Mobile Devices: Exploring the Role of Appropriation and Coping Strategies. *Research in Learning Technology, 15*(3), 247-258.

Ministry of Education. (2006). *ICT Strategic Framework for Education.*: The Ministry of Education, on behalf of the education sector agencies and the National Library of New Zealand.

Parr, M., & Ward, L. (2004). *Evaluation of the digital oppertunities project FarNet: Learning Communities in the Far North*. Wellington.

Prestridge, S. (2007). Engaging with the transforming possibilities of ICI: A discussion paper. *Australian Educational Computing, 22*(2), 3-9.

Richey, R. C. (1998). The Pursuit of Useable Knowledge in Instructional Technology. *Educational Technology Research and Development, 46*(4), 7-22.

Rivers, J., & Rivers, L. (2004). *A Summary of the Key Findings of the Evaluations of the Digital Opportunities Pilot Projects (2001 - 2003)*.

Salomon, G. (1993) *On the nature of pedagogic computer tools: the case of the Writing Partner* (pp. 179-196). Hillsdale, NJ: Lawrence Erlbaum Associates.

Van Dijk, J. (2006). Digital divide research, achievements and shortcomings. *Poetics, 34*(4-5), 221-235.

Wei, K. K., Teo, H. H., Chan, H. C., & Tan, B. C. Y. (2011). Conceptualizing and testing a social cognitive model of the digital divide. *Information Systems Research, 22*(1), 170-187.

Winter, M. (2004). *Digital Opportunities Pilot Project (2001-2003) Evaluation of Generation XP*. Wellington.

Yin, R. K. (2003). *Case Study Research: Design and Methods* (3rd ed.). Newbury Park: Sage Publications.

## Copyright